# Hardware Accelerators for Autonomous Cars: A Review

Ruba Islayem, Fatima Alhosani, Raghad Hashem, Afra Alzaabi, Mahmoud Meribout

*Abstract*— Autonomous Vehicles (AVs) redefine transportation with sophisticated technology, integrating sensors, cameras, and intricate algorithms. Implementing machine learning in AV perception demands robust hardware accelerators to achieve real-time performance at reasonable power consumption and footprint. Lot of research and development efforts using different technologies are still being conducted to achieve the goal of getting a fully AV and some cars manufactures offer commercially available systems. Unfortunately, they still lack reliability because of the repeated accidents they have encountered such as the recent one which happened in California and for which the Cruise company had its license suspended by the state of California for an undetermined period [1]. This paper critically reviews the most recent findings of machine vision systems used in AVs from both hardware and algorithmic points of view. It discusses the technologies used in commercial cars with their pros and cons and suggests possible ways forward. Thus, the paper can be a tangible reference for researchers who have the opportunity to get involved in designing machine vision systems targeting AV.

*Index Terms*— ADAS, ASIC, CNNs, CPU, Datasets, FPGA, GPU, Hardware Accelerators, SSD, Object Detection, YOLO

I. INTRODUCTION

AVs represent a groundbreaking technological innovation with profound implications for the field of transportation and beyond. Using a combination of sensors, cameras, lidar (Light Detecting and Range Technology), radar, and complex software algorithms, autonomous cars can observe their surroundings, make quick decisions in real time, and travel safely without the need for a driver. AVs have garnered significant interest recently and they hold a crucial place in transportation not just for the convenience they offer in relieving drivers but also for their capacity to revolutionize the entire transportation ecosystem. As per the WHO, approximately 1.3 million lives are lost each year due to road traffic accidents [2], and 94% of these accidents are because of human errors and distracted driving [3]. Therefore, their significance is underscored by their role in enhancing road safety by eliminating human errors, optimizing traffic flow, reducing congestion, and minimizing environmental impact [4]. Additionally, they offer increased mobility for individuals who cannot drive due to elderly or disabilities, promising a future that is safer, more efficient, and more accessible.

In recent decades, Machine Learning (ML) algorithms have played a pivotal role in advancing AV technology, particularly in the perception system. These algorithms facilitate the assessment of the vehicle's surroundings and identification of objects like pedestrians, vehicles, and traffic signals. The control system module utilizes this information to implement essential measures, covering actions related to braking, speed, lane changes, or steering adjustments [5]. The integration of artificial intelligence (AI) and ML is widespread in AV development, led by companies such as Waymo, Uber, and Tesla. This shift replaces conventional systems, reducing reliance on costly equipment like LRF (Laser Range Finder), LiDAR, and GPS [5]. Ongoing research aims to ensure AV safety by addressing challenges in modelling human-like driving behaviour for passenger comfort. ML, especially through the application of Convolutional Neural Networks (CNNs), assumes a central role in performing vital computer vision tasks essential for AV autonomy [5].

AVs leverage not just machine vision algorithms but also depend on hardware accelerators to furnish robust parallel computing frameworks, essential for managing the intricate responsibilities of perception, decision-making, and control [6]. These hardware accelerators encompass graphics processing units (GPUs), Central processing units (CPUs), Field-Programmable Gate Arrays (FPGAs), and Application-Specific Integrated Circuits (ASICs). The selection of these hardware accelerators for AVs can fluctuate based on several factors, including the AVs autonomy level, sensor configuration, computational demands, and safety prerequisites.

This review paper makes a valuable contribution to the field of AVs in several ways. Firstly, it addresses the absence of comprehensive review papers that discuss commercially available machine vision systems for autonomous vehicles. This serves as a valuable resource for researchers and industry professionals seeking insights into the practical implementation and industry relevance of these systems. Furthermore, this paper is groundbreaking because it covers all aspects of hardware accelerators and machine vision systems for AVs in one comprehensive document. Additionally, it tackles the issue of fragmented information by consolidating and presenting it in one accessible resource, making research and knowledge exchange more efficient. These contributions aim to advance research and drive progress in the development of AVs technologies.

The remaining sections of this paper are organized as follows:
- Section II provides a background overview of the hardware accelerators, sensors and machine vision algorithms commonly employed in AVs.
- Section III is dedicated to providing an overview of the machine vision algorithms used in AVs, offering insights into deep learning algorithms and machine learning algorithms used to detect relevant objects on the road.
- In Section IV, we conduct an in-depth exploration of some of the state-of-the-art processors and other potential hardware accelerators utilized in AVs to enhance machine vision algorithms.
- In Section V, we offer our conclusions, summarizing the main findings and implications presented throughout this paper.

## II. BACKGROUND

### A. Levels of ADAS

In 2014, SAE International introduced the J3016 standard, known as "Levels of Driving Automation" [7]. This standard classifies the Advanced Driver-Assistance System (ADAS) into six distinct levels of driving automation, as depicted in Fig 1 [7]. It commences at SAE level 0, where the driver maintains full control, and advances to SAE level 5, where vehicles achieve complete autonomy and handle all dynamic driving tasks without human intervention. In a level 5 system, the vehicle assumes full responsibility, even in the event of faults, errors, or accidents [8]. To reach higher autonomy levels, AVs depend on a combination of sensors and software to perceive their environment and navigate autonomously [9]. Currently, automotive manufacturers like Audi (Volkswagen) and Tesla have adopted SAE level 2 automation standards in the development of automation features like Tesla's Autopilot and Audi A80s Traffic Jam Pilot. In contrast, Alphabet's Waymo has been exploring a business model centered on SAE level 4 self-driving taxi services since 2016, offering rides within specific areas in Arizona, USA [7].

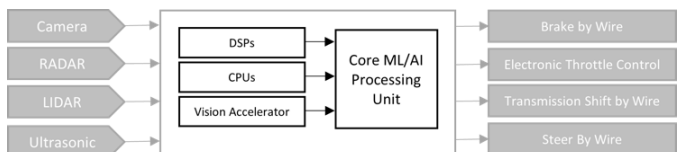

*Figure 1. An overview of the levels of driving automation* [7]

### B. General Structure of ADAS Systems

ADAS is a system that helps automobile drivers navigate and park without automating the whole process by employing camera-based sensors. It aims to minimize human accidents by processing important data about traffic, congestion levels, and road closures, among other things.

The brain of most ADAS systems is a hardware accelerator to perceive the car's surroundings to avert danger. They typically comprise four perception sensors LIDAR, RADAR, Cameras, and Ultrasonic Sensors [10]. The data from these sensors is processed using a dedicated hardware accelerator and fused together to identify nearby objects such as pedestrians, vehicles, lanes, and traffic signs [11]. Finally, the pre-processed data is explored by other components such as the brake, steering, and throttle control to react accordingly based on the obstacles faced. The entire process is depicted in the figure below.

*Figure 2. ADAS General Processing Structure*

### C. Perception Sensors Used by Manufacturers

AVs utilize 4 main types of perception sensors: cameras, RADARs, LIDARs and ultrasonic sensors. The cameras which arguably yield the most useful and larger information may be of different type: fish-eye cameras for wide-angle coverage, monocular cameras for basic visual data, stereo cameras for depth perception, and 360-degree cameras for panoramic views [7]. Also, depending on their focal length and orientation, they can be used to cover different views surrounding the car: near/far front-view, side-view, rear-view, surround-view, and built-in cameras, based on the different applications and scenarios [12].

RADARs, Radio Detection and Ranging sensors, detect and locate objects within a specific range from the car. Most AVs employ 3 variants of RADARS: long-range, medium range and short-range [13].

LIDAR, sensors use laser beams to detect and measure the time it takes for the beams to reach the object; thereby allowing the system to create a 3D map of the environment [14]. Their high accuracy, along with their effectiveness in low-light conditions, makes them an important component for autonomous vehicles.

Ultrasonic sensors provide short-distance data and are typically used for parking assistance and backup warning systems, as far as there is no rain [9]. However, unlike cameras, they can operate in foggy and dusty weather conditions.

Among all sensors, the camera is the main visual sensor of the ADAS system due to its ability to perform high-resolution tasks, including classification and scene understanding that require color perception. There has been a growing belief among researchers and even companies that autonomous driving will be possible with cameras only. Tesla is one of such companies as it uses AI and dedicated hardware accelerators to process video data in order to simultaneously estimate the depth, velocity, and acceleration using camera input [15]. However, this system has yet to demonstrate its reliability as some fatal accidents have occurred since its adoption [16]. In this paper, the focus will be made on such camera-based systems as they have great potential to achieve the SAE level 5 in the near future. The continuous advance of AI and associated hardware accelerators is the main catalyzer for this optimism. Fig 3 shows the overall sensor placement in AVs.

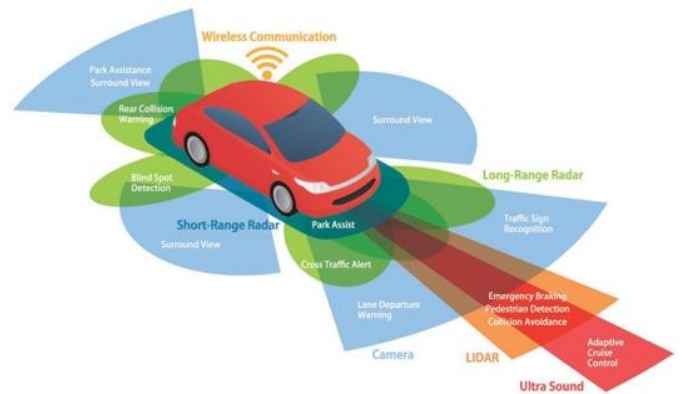

*Figure 3. Typical Placement of Sensors Around an Autonomous Vehicle* [18]

*D. Need for Hardware Accelerators*

In AVs, traditional computer processors, such as Intel-7 CPU, lack the power to host computationally intensive machine vision algorithms hence the need for special-purpose coprocessors or AI accelerators such as GPUs, FPGAs, and ASICs have been widely used in automobiles as shown in Fig 4 [17-19].

On one end, while multicore CPUs have general flexibility and can execute AI and machine learning workloads, there is no specific support for them, and they are not energy efficient. On the other hand, GPUs, which are also versatile, feature higher levels of parallelism using even single and double precision arithmetic. Thus, they are more adequate to handle memory intensive tasks required in machine vision algorithms to yield higher throughput than multicore CPUs [6]. FPGAs also offer adaptability for customizing parallelism, data types, and hardware architecture to suit specific applications. They are useful for accommodating lighter versions of modern DNN models featuring quantized weights and reduced number of layers [6]. Furthermore, ASICs, which are customized hardware chips designed for specific applications, offer high performance and efficiency tailored to their designated functions in terms of execution time and power consumption [6]. However, they lack flexibility and are designed for specific purposes. Therefore, systems designers must consider a blend of processor resources to meet their application needs. Tesla, NVIDIA, Qualcomm, and Mobileye, among other companies, have been working on developing their own AI accelerators targeting AV applications.

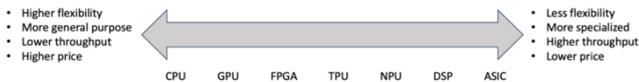

*Figure 4. Spectrum of Hardware Accelerators* [17]

*E. Machine Vision and emergence of CNNs*

One of the most significant advancements in machine vision technology is the integration of CNNs. They are one of the best machine learning algorithms for recognizing image content and have demonstrated good performance in image segmentation, classification, detection, and retrieval related tasks [20]. Some of the most widely used CNN models in machine vision targeting AV are YOLO, Faster RCNN, SSD, and MobileNet [21]. These models are well-known for their exceptional performance in various image-related tasks, making them essential tools in the field of AVs.

*F. State of the Art AVs and their level of autonomy*

Car manufacturers have been researching AVs since the 1920s [22]. The first modern AV in 1984 had level 1 autonomy, followed by a level 2 AV from Mercedes-Benz in 1987 that could control steering and acceleration with limited human supervision [23]. In 2014, Tesla became the pioneer in bringing AVs to the commercial market with their Autopilot system, offering level 2 autonomy [23]. Tesla's AVs heavily rely on sensors for self-navigation and decision-making, including a suite of six forward-facing cameras and ultrasonic sensors [24]. Volvo, in 2017, introduced their Drive Me feature, providing level 2 autonomy, allowing their vehicles to travel autonomously in specific weather conditions [25]. Furthermore, Waymo launched a driverless taxi service with level 4 autonomy in 2018 in the Phoenix area, USA, serving 1,000 to 2,000 riders weekly, with 5-10% of these rides being entirely autonomous [26]. Cruise Automation, in 2017, began testing a fleet of 30 vehicles with level 4 autonomy and introduced their self-driving Robotaxi service in 2021 [23]. Cruise AVs utilize a sensor cluster, featuring a front radar and cameras along with lidar sensors mounted on top to offer a comprehensive 360-degree view of their surroundings [26]. However, it's worth noting that the California Department of Motor Vehicles (DMV) has recently revoked Cruise's permits for testing and operating fully autonomous vehicles on the state's roads due to several reasons, including their failure to disclose information about a pedestrian accident, where a Cruise vehicle struck a pedestrian and dragged them along the road [27]. Although both Waymo and Cruise aspire to achieve level 5 autonomy, their AVs are presently classified as level 4 due to the absence of a guarantee for safe operation in all weather and environmental conditions as well as to the road traffic accidents they have caused.

*G. Challenges in AVs*

Recent research in machine vision for AVs has achieved significant progress but faces various challenges that warrant further investigation. Firstly, real-time object detection is complex due to the need for simultaneously processing several video streams in real-time (more than 10 video streams corresponding to different orientations and zooming of the cameras in most of the cases) [23]. A limited number of studies, such as [28] and [29], considered multi-frame perception, which uses data from previous and current time instances. Moreover, semi-supervised object detection, involving annotated data for model training, faces challenges in annotating diverse scenarios, which are essential for the models' adaptability in real-world AV driving scenarios [23]. Recent research [30-32] recommends semi-supervised transformer models for improved accuracy but deploying them on embedded onboard computers poses memory challenges requiring further investigation. Finally, object detector performance varies with changing environmental conditions like light and weather. Addressing this issue involves collecting diverse weather data, crucial for training reliable object detectors. The Waymo open dataset offers such diversity to improve detector performance [33]. Such open datasets are vital for ensuring consistent performance in various environmental conditions in AVs. The other challenge is the increasing complexity of the hardware accelerators which require an in-depth hardware skill as well as masterminding of both the AI algorithms and the associated firmware and the real-time operating system structure. It is rare to have these attributes featured by one single researcher which require multidisciplinary teamwork.

III. MACHINE VISION ALGORITHMS FOR AVS

In the past, the computational capabilities of hardware accelerators were not powerful enough to support the integration of CNN models. Most of the traditional vision algorithms were not using CNN models, primarily because they are computationally intensive. Nevertheless, with the advancement of hardware accelerators, their implementation at reasonable power consumption to be performed in real-time is becoming possible. As a result, CNN models have replaced most traditional image processing methods. Moreover, these image processing methods are not reliable because they rely on manual feature engineering, making them less adaptive and time-consuming, especially for complex object detection tasks. They often struggle to recognize objects in diverse driving scenarios, requiring frequent adjustments for changes in object scale, rotation, and varying environmental conditions, which can limit their reliability and effectiveness. This is, in fact, a big drawback of traditional legacy image-processing algorithms dedicated to autonomous cars. The challenge lies in their ability to accurately detect vehicles, where even minor alterations in a vehicle's appearance can lead to detection failures. An illustrative example is the disruption caused by an extended arm from a car's window, resulting in a system malfunction. In contrast, CNN models prove advantageous as they exhibit robust performance, making them the preferred choice.

The illustration in Fig 5 outlines the autonomous vehicle processing pipeline employed in today's machine vision systems. The pipeline, structured in discrete stages, facilitates the seamless flow of information from sensor data to high-level decision-making. Specialized CNN models tailored for distinct object detection tasks enhance vehicle safety and overall performance. Beginning with the camera capturing images, the pipeline includes video decoding for bandwidth optimization, image preprocessing for tasks like resizing and noise reduction, and specialized models for detecting vehicles, pedestrians, lanes, and traffic signs. The high-level preprocessing phase integrates these outputs to make informed decisions, addressing tasks such as safe distance calculation and responding to traffic signs. Finally, the decoder translates processed data for visualization, control, and output, including displaying object detections on a user interface and transmitting commands to vehicle actuators.

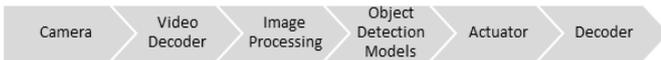

*Figure 5. AVs Processing Pipeline*

### A. Object Detection Algorithms

Object detection comprises two key tasks: localization, determining the precise object position in an image or video frame, and classification, assigning a specific class to the object. This classification can include identifying objects like pedestrians, vehicles, or traffic lights [23]. Detection and classification can be done in a single (e.g. R-CNN) or two independent stages (e.g. YOLO) [34]. Unlike two-stage detectors, which rely on a separate region proposal step for bounding-box prediction, one-stage detectors perform this directly from input images, resulting in faster performance [34].

#### 1) Two-Stage Detectors
##### a) R-CNN

R-CNN is a two-stage object detection framework that converts the traditional object detection problem into a feature acquisition problem for regions and a classification problem for proposals [35]. To minimize information loss and enhance efficiency, spatial pyramid pooling (SPPNet) is used for feature extraction, providing features of various sizes [35]. R-CNN has been demonstrated to yield high performance for AVs, specifically for detecting various objects, including pedestrians, cars, and traffic signs [36]. Even though R-CNN achieves cutting-edge results, it is very slow to train and test due to the need to process thousands of regional proposals for each image [36].

In the initial phase of the R-CNN methodology, as shown in Fig 6, approximately 2,000 region proposals are generated to encompass potential objects [34]. Then, each region goes through a backbone network such as AlexNet, to extract feature representations consisting of 4,096 dimensions. To enhance the accuracy of object classification, the system uses a Support Vector Machine (SVM) for making predictions. Furthermore, the system utilizes Fully Connected Layers (FCLs) to refine these predictions. Adjustments to the bounding boxes are made more precise with a Bounding-Box regression technique and a method called greedy non-maximum suppression (NMS). By following this process, R-CNN achieved a mean average precision (mAP) of 58.5% on the Pascal VOC dataset.

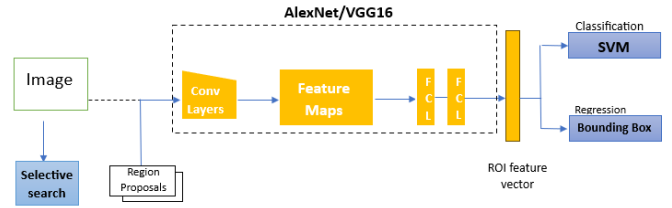

*Figure 6. R-CNN Process Pipeline*

##### b) Fast R-CNN

Fast R-CNN model enhances object detection by analyzing the entire image simultaneously, making it faster and more accurate than the previous R-CNN model. As shown in Fig 7, it begins by processing the image through a CNN to create a feature map. Regions of interest (ROIs) are then identified on this map, and through ROI pooling, fixed-size feature vectors are generated. These vectors are employed in FCLs for predictions, using 'softmax' and 'bounding-box regression' for categorization and precise location determination, respectively. It achieved mAPs of 70.0%, 68.8%, and 68.4% on Pascal VOC 2007, 2010, and 2012 datasets when trained with VGG-16 [34]. However, it relies on external region proposals, which is computationally expensive [6], therefore, Faster R-CNN was introduced.

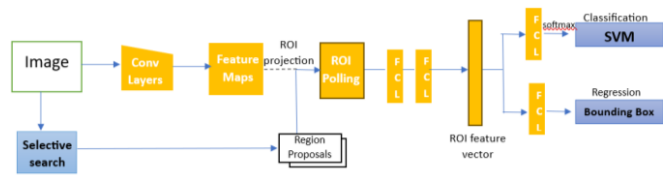

*Figure 7. Fast R-CNN Process Pipeline*

##### c) Faster R-CNN

Faster R-CNN builds upon the improvements made by R-CNN and Fast R-CNN by eliminating the need for selective search and introducing a Region Proposal Network (RPN) [23]. As shown in Fig 8, this small convolutional network generates region proposals directly from the CNN's feature map, streamlining the process of extracting bounding boxes and significantly enhancing training and computing speed. Moreover, Faster R-CNN employs a separate network to feed the ROI to the ROI pooling layer and the feature map [34]. These inputs are subsequently reshaped and utilized for prediction. In Faster R-CNN, the number of ROIs is not a constant value and is defined by the size of the feature map. Thus, the region proposals were implemented on GPUs with nearly free computation cost compared to previous baselines [34]. This optimized architecture allows Faster R-CNN to achieve a rapid 6 frames per second (FPS) inference speed on a GPU while maintaining state-of-the-art detection accuracy on Pascal-VOC 2007 [6]. Despite speed and accuracy improvements, the two-stage approach still falls short of real-time performance requirements.

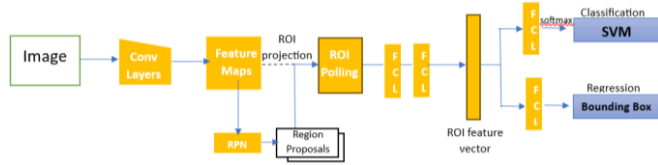

Figure 9. Faster R-CNN Process Pipeline

*2) Single-Stage Detectors*

*a) YOLO*

Although Faster R-CNN reduces region proposal overlaps, it still hinders performance due to repeated calculations [34]. A new hybrid CNN-based architecture called YOLO (You Only Look Once) addresses this issue. It can predict objects with a single pass and efficiently handle object identification and classification by combining region proposals and detection into one stage [36]. The architecture of YOLO models is illustrated in Fig 9. It consists of three components: the backbone network, the neck, and the head. The backbone network is a convolutional neural network that extracts features from the input image. The neck consists of a series of convolutional layers that combine the features from the backbone network to form a high-level representation. Lastly, the head is composed of convolutional layers that generate the final predictions of bounding boxes and class probabilities [37]. Fig 10 illustrates the process pipeline of YOLO models, they divide input images into a set of grid cells, with each cell responsible for predicting bounding boxes and class probabilities for the objects present [34].

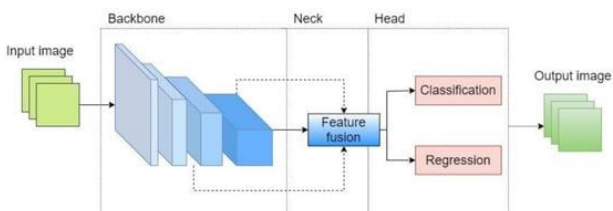

Figure 10. YOLO Architecture [121]

Previous YOLO models faced limitations in detecting small objects, generally with varying object aspect ratios, and issues with their loss functions [34]. To address these limitations, improved versions of YOLO were suggested. YOLOv8, the latest version of YOLO excels in precision and speed, making it ideal for detecting small objects using advanced techniques such as bounding boxes, multi-scale prediction, and feature fusion [38]. YOLOv8 introduces five different versions (nano, small, medium, large, and extra-large) and supports various vision tasks, including object detection, segmentation, pose estimation, tracking, and classification. It utilizes a modified backbone called the C2f module, which combines high-level features with contextual information and employs an anchor-free model with a decoupled head to enhance overall accuracy [39]. In the output layer, the model employs the sigmoid activation function to determine the abjectness score, indicating the probability that the bounding box contains an object. Additionally, the softmax function is utilized for class probabilities, indicating the objects' probabilities belonging to each possible class. When evaluated on the MS COCO dataset test-dev 2017, YOLOv8x achieved an Average Precision (AP) of 53.9% with a 640-pixel image size and a speed of 280 FPS on an NVIDIA A100 with TensorRT [39].

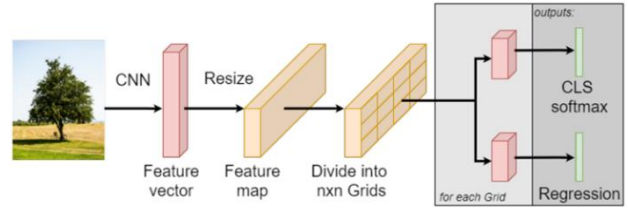

Figure 8. YOLO Process Pipeline [121]

*b) SSD*

Single-Shot Detector (SSD) models offer another good alternative for real-time video applications as they efficiently handle both classification and localization tasks on the entire image, ensuring accuracy [23]. The SSD model is structured with six stages in a hierarchical design to form a single forward pass network [34]. The goal is to achieve hierarchical feature extraction, where each layer in the hierarchy contributes to object classification and bounding-box detection with different levels of semantic information (Fig 11). To optimize efficiency, each stage incorporates a fast non-maximum suppression (NMS) technique, removing redundant bounding boxes in post-processing.

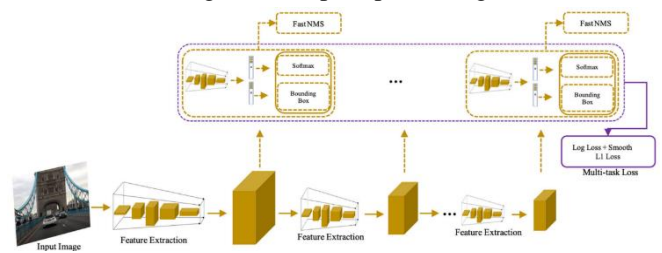

Figure 11. SSD Process Pipeline [34]

Table I provides a performance comparison of different object detection models for AVs. The models are evaluated based on key metrics, including model size, FPS, and Mean Average Precision (mAP). Notably, the table shows that studies on autonomous driving face limitations due to the trade-off between accuracy and real-time operation speed, restricting the applicability of self-driving systems. For instance, while some models exhibit high accuracy, they may compromise on operational speed, and vice versa. YOLOv8 stands out as a state-of-the-art model, achieving an impressive mAP of 95.1. Its model size of 9.2 MB and FPS of 221 showcase a well-balanced performance, ideal for accurate and efficient object detection tasks without compromising on real-time processing speed. Additionally, SSD impressively achieves a high mAP of 90.56, showcasing its proficiency in accurately identifying objects. Although its specific model size is not disclosed, SSD exhibits a remarkable FPS rate of 105.14, indicating rapid real-time detection capabilities.

TABLE I
PERFORMANCE METRICS OF OBJECT DETECTION MODELS

| Model | Dataset | Hardware Platform | Model Size (MB) | FPS | mAP |
|---|---|---|---|---|---|
| DYNAMIC R-CNN [34] | MS COCO | GeForce RTX 2080TI | 550 | 13.9 | 49.2 |
| YOLOv5x [40] | VOC2007 + 2012 COCO | GeForce GTX 1650 | 87.37 | 10.09 | 81.18 |
| MobileNet-YOLO [40] | VOC2007 + 2012 COCO | GeForce GTX 1650 | 3.23 | 73.39 | 73.17 |
| YOLOv7-tiny [41] | TIB-Net | GeForce RTX 3070 | 12.2 | 227 | 85 |
| YOLOv8 [41] | TIB-Net | GeForce RTX 3070 | 9.2 | 221 | 95.1 |
| SSD [42] | PASCAL VOC 2007 + 2012 COCO | GeForce RTX 2080TI | - | 105.14 | 90.56 |
| Faster R-CNN (VGG16) [43] | PASCAL VOC 2007 + 2012 COCO | CPU | - | 7 | 73.2 |
| Fast R-CNN [43] | PASCAL VOC 2007 + 2012 COCO | CPU | - | 0.5 | 70.0 |

B. *Algorithms for Detected Objects in AVs*

  1) *Lane Detection*

Lane detection algorithms rely on line detection and edge detection [44]. Initially, traditional image processing algorithms were used such as the Hough transform which is one of the widely used algorithms as it features high level of parallelism and accuracy of detection [45]. Other image processing-based algorithms including LaneATT [46], RANSAC, control point detection, lane marking clustering and fan-scanning line detection, were also employed [44], [46]. With the development of deep learning techniques, CNN algorithms such as CNN, RNN, R-CNN and YOLO family have been used for lane detection [44], [46]. According to [47], CNN models reported a 90% accuracy for lane detection as compared to traditional image processing algorithms, which have an accuracy of 80% [44]. Caltech Lane, KITTI, TuSimple, and CuLane are the most used datasets to train algorithms for lane detections [44], [46].

  2) *Pedestrian Detection*

In the past, traditional object detection algorithms such as VJ detector and Histogram of Oriented Gradients (HOG) have been used for pedestrian detection, all of which provided high accuracy rates [48]. In 2008, the Deformable Parts Models (DPM) detection algorithm was proposed [49]. DPM divided pedestrians into different parts and then treated them as a collection consisting of different parts during object classification. At that time, the algorithm had the best detection results until the optimization methods using deep learning emerged. RFCN, Mask RCNN, RetinaNet, YOLO, and SSD are commonly used algorithms for pedestrian detection [50], [51]. Additionally, CompACT, SAF RCNN, and ALFNet are proposed optimized algorithms specific for pedestrian detection tasks [51-53]. In [54], YOLO-R, an optimized YOLO algorithm has been proposed, which has a high precision of 98.6%. In comparison, R-CNN models typically reported a precision ranging from 70-80% [55], [56]. To train pedestrian detection algorithms, Caltech, KITTI, CityPersons, EuroCity, INRIA and COCO are among the most used datasets [57].

  3) *Traffic Sign Detection*

Traffic sign detection algorithms are essential in analyzing, detecting, and categorizing traffic signs based on their shape, color and drawings on them [58]. Traffic sign algorithms are classified into two types: machine learning based, and deep learning based. Machine learning based algorithms include Support Vector Machine (SVM), and AdaBoost to detect traffic signs accurately using handcrafted features [58]. On the other hand, deep learning algorithms such as CNNs and RNNs have been more commonly used recently due to their ability to automatically learn complex features from raw data, reducing the need for manual extraction [58]. For instance, enhanced algorithms based on ResNet and CNN, as introduced by [59], demonstrate effective capture of intricate features in traffic signs. Utilizing the Kaggle traffic sign dataset, the ResNet-based model achieved an impressive recognition accuracy of 99%, while the CNN-based model attained a recognition accuracy of 98%. GTSRB, COCO and TT100K are some of the most used datasets to train traffic sign detection algorithms [58], [60], [61].

  4) *Traffic Light Detection*

Traditional image-processing traffic light algorithms can be processed into two steps: feature extraction and template-matching [62]. Feature-extraction algorithms are used to know the features of the traffic light signal, and commonly used algorithms are SIFT, PCA-SIFT, and SURF [63-65]. On the other hand, template-matching algorithms, or classifiers are used to match and classify features. Adaboost, SVM, and LDA are some of the algorithms used for template matching [68-70]. While these algorithms are still being used, they lack generality where even a marginal change in the object appearance would cause false negatives. With the development of deep learning algorithms, the YOLO family and RCNN series have been widely used for traffic light detection [62]. Most of the recent research has been focused on optimizing YOLO algorithms [69-73]. Most notably, the most recent version of YOLO, YOLOv8 has been optimized for traffic light detection in [72], achieving a high mean average precision of 98.5% as compared to the implementation of Faster R-CNN in [74], which achieves a maximum mean average precision of 86.4%. In order to train the algorithms, LISA, Bosch, and DriveU are some of the main datasets created specifically for traffic light color detection [75-77].

IV. HARDWARE ACCELERATORS

Recent advancements in computer vision algorithms have been primarily driven by deep learning and the availability of extensive datasets. Hardware acceleration has played a significant role in this progress, providing parallel computing architectures that enable the efficient training and execution of complex neural networks. State-of-the-art processors such as the ones manufactured by Tesla, NVIDIA, Mobileye, and Qualcomm hardware accelerators have been among the most widely used accelerators in the industry to power autonomous vehicles. However, FPGAs and TPUs are also other hardware accelerators that hold great protentional to be used to other autonomous vehicles. When hardware accelerators are combined properly and optimized, they can make-up for the drawbacks in each other, paving the path for attractive heterogenous hardware solution. In this section of the report, an overview is first given of the different state-of-the-art processors used in AVs, concluding it with a comparison between them.

*A. State-of-the-Art Processors Targeting AVs*

In the fast-changing world of AVs, the core of advanced technology lies in state-of-the-art processors. Companies like NVIDIA, Tesla, Qualcomm, and MobilEye lead the way in shaping the intelligence and effectiveness of self-driving systems using their own hardware accelerator. Besides NVIDIA, all other manufactures do not commercialize their respective processors, which may alter their progress in both the software and hardware areas. This has led NVIDIA to lead the race by offering cutting edge processors effectively used not only in AVs but also in other related areas such as generative AI, metaverse, and robotics. Indeed, most of the algorithms dedicated for AVs were developed on NVIDA platforms. This section explores the details of these powerful processors, exploring their unique features, innovations, and contributions to improving self-driving technology.

*1) TESLA*

In 2019, Tesla introduced Hardware 3.0 (HW3), its dedicated AI self-driving hardware supporting Full Self-Driving (FSD) technology [78]. This custom-designed chip is built on Samsung's 14 nm process [79]. As shown in Fig 12, it integrates 3 quad-core Cortex-A72 clusters, totalling 12 ARM Cortex-A72 CPUs operating at 2.2 GHz, 2 neural processing units (NPUs) operating at 2 GHz, achieving a peak performance of 36.86 TOPS, and a GPU operating at 1 GHz with a capacity of 600 GFLOPS [79]. The FSD chip also features an image signal processor (ISP) for handling the eight High Dynamic Range (HDR) sensors, H.265 video encoder, and camera serial interface (CSI) for managing sensors, along with a conventional memory subsystem supporting 128-bit LPDDR4 memory at 2133 MHz [80]. The system features two independent FSD chips, each with its dedicated storage and operating system [78]. In case of a primary chip failure, the backup unit seamlessly takes over. Notably, the HW3 outperforms the previous NVIDIA DRIVE PX 2 AI platform, delivering 36.86 TOPS compared to the previous 21 TOPS [78]. The FSD computer consumes 72 Watts, with 15 Watts attributed to the NPUs [80].

Various object detection algorithms are employed in Tesla cars to recognize and monitor objects within the visual scope of a vehicle. This includes conventional computer vision methods like HOG or employ more sophisticated deep learning methodologies such as YOLO and R-CNN [81].

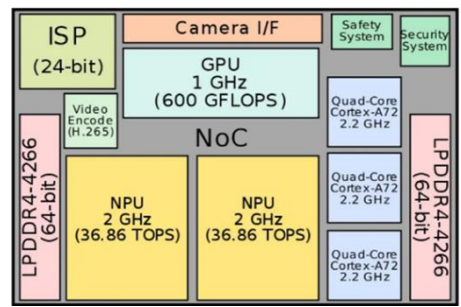

*Figure 12. FSD Block Diagram* [79]

*2) NVIDIA*

Nvidia Jetson is a low-power computing board series, integrating an ARM architecture CPU used to accelerate machine learning applications using tensor cores [82]. Most notably, Jetson Xavier, Jetson Nano, and Jetson Orin have been used for autonomous vehicle applications.

The NVIDIA Jetson AGX Orin, released in 2023, is programmable using CUDA and Tensor APIs and libraries, offering 275 TOPS with power configurable between 15W and 60W [83]. Jetson AGX Orin modules feature the NVIDIA Orin SoC, which is built on an 8nm chip, with a NVIDIA Ampere architecture GPU, Arm Cortex-A78AE CPU, next-generation deep learning and vision accelerators, and a H.264/5 video encoder and a video decoder. Furthermore, it supports LPDDR5 memory and has a DRAM capacity of 32GB or 64GB.

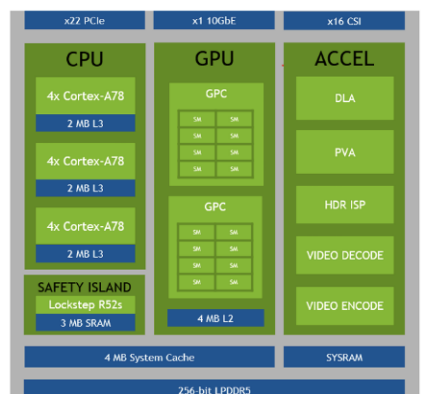

*Figure 13. NVIDIA Jetson Orin AGX Block Diagram* [122]

In addition to being very powerful, the other main advantage of this processor is its wide availability for researchers, and to feature a powerful software development kit. Thus, Yassin K. et. al [84] proposed a lane detection algorithm based on CNN Encoder–Decoder and Long Short-Term Memory (LSTM) networks, implemented on the NVIDIA Jetson Xavier. Notably, it achieves a frame rate of 6.78 FPS and takes 147 ms to process a 1280*720 input image as compared to Intel Core i7-2630QM CPU processor, which achieves a frame rate of only 3.62 FPS and an execution time of 276 ms. In [85], LW-YOLOv4-tiny is implemented on the Nvidia Jetson Nano for rapid object detection and it achieves an execution speed of 56.1 FPS.

Automotive manufacturers like Audi, Mercedes-Benz, and Volvo partnered with Nvidia to incorporate NVIDIA Jetson into their autonomous vehicles, aiming to achieve advanced self-driving capabilities [86-88].

### 3) Qualcomm Snapdragon

In January 2022, Qualcomm launched the Snapdragon Ride Vision System, employing cutting-edge 4-nanometer processing technology in a flexible and scalable vision software stack [89]. Integrated with the proven Vision Stack, it enhances front and surround-view cameras for ADAS and automated driving [89]. The Snapdragon Ride SoC, a key element of the hardware platform, is tailored for ADAS needs, featuring machine learning processors, image signal processors, vision and graphics acceleration, dedicated DSPs, GPU technology, multi-core ARM-based CPU, and safety and security systems [90]. With excellent thermal efficiency, it delivers 30 TOPS for L1/L2 applications and over 700 TOPS at 130W for L4/L5 autonomous driving [91].

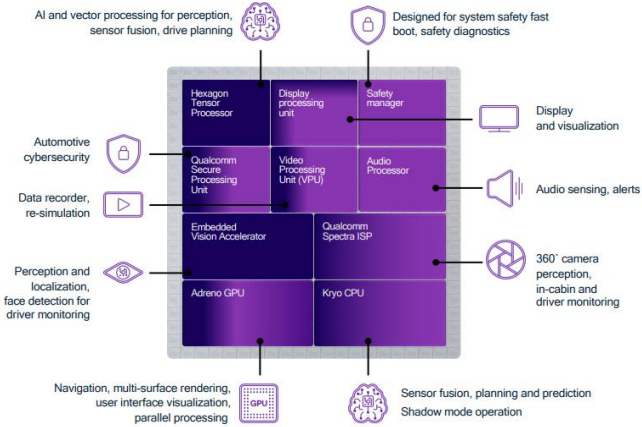

*Figure 14. Qualcomm Snapdragon SoC Architecture* [123]

The Platform is designed to serve three distinct segments of autonomous systems [90]. In the Active Safety ADAS segment, it addresses functions like autonomous braking, traffic sign recognition, and lane assist, employing a passively cooled ADAS chip delivering 30 TOPS. The Convenience ADAS segment encompasses applications such as self-parking, highway driving, and urban driving, utilizing a combination of SoCs with the goal of delivering 60 to 125 TOPS. Lastly, the Fully Autonomous Driving segment is tailored for autonomous urban and highway driving, employing two ADAS chips and one to two ML accelerators, capable of delivering up to 700 TOPS at 130W. However, there is no information disclosed about the specific types of machine learning algorithms used to employ these segments.

Qualcomm's advanced processors are favoured by top AV companies like Waymo, Cruise, and Argo AI for their high performance and efficiency. Qualcomm also leads a collaboration with BMW to develop a comprehensive range of driving features [92]. This includes advanced image recognition utilizing front, rear, and ambient cameras, supported by a dedicated processor (Computer Vision SoC) and a high-performance computing unit tailored for ADAS [92].

### 4) Mobileye EyeQ

The Mobileye EyeQ system-on-chip is a chip using a single camera sensor to provide passive and active autonomous driving. In mid-2023, EyeQ6 became Mobileye's main ADAS SoC, coming in two variants: EyeQL, which is the entry-level chip powering forward-facing camera systems, and EyeQH, which is more full-featured and has multiple surrounding cameras [93]. It features a CPU with MIPS architecture, featuring multithreading. Compared to its predecessors, it features two important GPUs, a small-scale ARM MALI GPU for AR image overlay, and the other GPU is unidentified; however, it is dedicated to handling OpenCL for stereo matching [94]. Furthermore, MobilEye launched EyeQ Ultra shown in Fig 14 [95], which is expected to power autonomous vehicles from 2025. EyeQ Ultra is built on a 5nm chip, has 12 CPU cores with 24 threads based on RISC-V architecture, a GPU, a vision processor, an image signal processing core, and 16 convolutional neural network clusters. Furthermore, it can encode videos of H.264/5 standard and it supports a memory of LPDDR5X. As compared to NVIDIA, which focuses on deep learning algorithms, Mobileye solutions still utilize convolutional computer vision algorithms aided by deep learning algorithms [96]. Some of those algorithms include True Redundancy for Sensor Fusion, Road Experience Management, and Intelligent Speed Assist [96], [97]. However, while the solutions are known to the public, most of the underlying CNN algorithms used by Mobileye remain undisclosed. Automative manufacturers such as Ford, NIO, Volkswagen, BMW and Nissan have collaborated with Mobileye to incorporate their EyeQ solution [98], [99].

A summary table of the hardware processors discussed, and their key features is shown in Table III. Most notably, the NVIDIA Jetson Orin AGX has the lowest power consumption. The Qualcomm Snapdragon SoC offers the highest peak performance of 700 TOPS coming at a cost of an extremely high-power consumption of 130 W. However, as much of the architecture of some of the chips is undisclosed such as the specific number of cores and their memory capacity, it is difficult to make a viable comparison between them all. Among all commercial accelerators for AVs, NVIDIA has the most open-source software platforms, making it the hardware processor of choice as it can be easily catered to the requirements of different AV manufacturers.

TABLE II
THE KEY DISTINCTIONS AMONG THE FOUR STATE-OF-THE-ART PROCESSORS

| Feature | TESLA FSD HW3 | Nvidia Orin SoC | Mobileye EyeQ Ultra | Qualcomm Snapdragon SoC |
|---|---|---|---|---|
| Chip Width | 14nm | 8nm | 5nm | 4nm |
| CPU Architecture | ARM Cortex-A72 (12 cores) | ARM Cortex-A78AE | 12 Cores, 24 Threads, ARM-based | Multi-Core ARM-based CPU |
| Video Encoding/ Decoding | Video Encoder (H.265) | Video Encoder and Decoder (H.264/5 and AVI) | Video Encoder (H.264/5) | Not specified |
| Memory Type and Bandwidth | LPDDR4 (2133 MHz, 68 GB/s) | LPDDR5 | LPDDR5X | Not specified |
| Peak Performance | 73.7 TOPS | Up to 275 TOPS | 176 TOPS | Over 700 TOPS |
| Power | 72 W | 15W to 60W | Under 100W | Up to 130W |
| Memory Capacity | 16GB RAM | 32GB/64GB DRAM | Not specified | Not specified |

## B. Other Hardware Accelerators

Beyond state-of-the-art processors based on GPU or CPU architectures, other hardware accelerators are crucial to advancing the capabilities of AVs. Specialized hardware accelerators like FPGAs and ASICs (specifically TPUs) have gained importance due to their ability to deliver lower latency and higher throughput compared to traditional general-purpose CPUs. As a result, there is a growing demand for these robust hardware accelerators in the industry. Manufacturers are actively incorporating them into their hardware solutions to meet the requirements of implementing high-performance algorithms and applications in AVs.

### 1) Field Programmable Arrays (FPGA)

FPGAs, comprise an array of configurable logic blocks and programmable interconnects, which can be tailored to create intricate digital circuits [100]. They also comprise hundreds of DSP blocks to handle multiply-and-accumulate intensive operations. They are specifically designed for executing fixed-point operations using a hardware-centric programming approach. In the field of autonomous vehicles, the utilization of FPGA-based systems aims to achieve two primary objectives: cost reduction in driverless technology and enhanced energy efficiency of their controllers. As a result, they offer substantial acceleration in image processing applications, rendering these systems significantly faster and more power efficient [101]. Their innate parallelism aligns seamlessly with the data-intensive demands of sensory fusion in autonomous vehicles.

Xilinx and Intel (Altera) have been at the forefront when it comes to manufacturing FPGAs for ADAS. Most notably, Xilinx's ZYNQ FPGA incorporates multiple ARM processors and leverages nested-loop algorithms to accelerate CNN inference [41]. It achieves an impressive 14 frames per watt (fps/watt) when handling CNN tasks, surpassing the Tesla K40 GPU, which achieves only 4 fps/watt [102]. Other XILINX boards recently used for CNNs include XILINX's Virtex-7 and Kintex-7. In 2021, XILINX released the Kria KV260, which is a development platform for Kria K26 System-On-Modules built for advanced machine vision application developments without requiring advanced hardware design knowledge [103]. Notably, it has a high number of DSPs and logic cells. However, this comes at the cost of higher power consumption.

Intel Cyclone 10 also exhibits great performance given the high number of logic elements and digital signal processing blocks, leveraging its parallel processing speed and flexibility. Table III provides a list of some of the most recent commercial FPGAs used for CNN algorithms. The FPGAs are compared based on the number of logic elements, DSPs, memory, and availability of video decoder, among other things. Generally, most recent review papers about CNN models use XILINX ZNYQ. Xilinx Virtex-7 shows superior performance, as indicated by the high number of logic elements and DSPs. However, its high versatility comes with a high cost. XILINX Kria KV260 and Intel Cyclone 10 GX are also FPGAs that show great potential for ADAS applications.

A study conducted by [110] proposes a reconfigurable CNN accelerator tested using YOLOv2-TINY and applied on XILINX KV260 FPGA, NVIDIA GeForce RTX2060 GPU and AMD Ryezen7 4800 H CPU. The KV260 board has the lowest operating frequency at 250 MHz and the lowest power consumption of 5.220W as compared to the GPU's power consumption of 175 W and the CPU's, 45W. When comparing the implementation of YOLOv2-TINY on the KV260 to the ZYNQ FPGA, the KV260 has a high data precision of 32 bits as compared to 16 bits on ZYNQ. Additionally, the KV260 has a high peak energy efficiency of 13.62 GOPS/W as compared to 6.3125 GOPS/W on the ZYNQ FPGA. With the increasing computational demands for AVs, FPGAs offer a great alternative solution to traditional processors given their high parallelism, low power consumption and high energy efficiency. Furthermore, FPGA manufacturers are moving towards creating FPGAs catered towards handling machine vision algorithms as exemplified by XILINX's KV260 FPGA.

### 2) Tensor Processing Unit (TPU)

Unlike more generic co-processors like GPUs and FPGAs, Google's TPUs, which are ASIC-based processors, are designed to meet specific requirements and are increasingly being adopted in the automotive industry [6].

TABLE III
SPECIFICATIONS OF FPGA BOARDS USED IN CNN ALGORITHMS

| FPGA Kit | System on Chip/Module | Logic Elements | DSP Blocks | Memory |
|---|---|---|---|---|
| **ALINX SoM AC7020: SoC Zynq7000 XC7Z020 Module** [104] | Zynq 7000 XC7Z020 SoC | 85K | 220 | 1 GB RAM DDR3L + 16 MB Quad-SPI Flash |
| **Avnet ULTRA96-V2 Development Board** [105] | Zynq Ultrascale+ MPSoC | 154K | 360 | Micron 2 GB (512M x32) LPDDR4 Memory |
| **Xilinx Kria KV260 Vision AI starter kit** [103] | Kria K260 SoM | 256K | 1.2K | 4 GB DDR4 |
| **Xilinx Kintex-7 KC705** [106] | Kintex 7 XC7K325T2FFG900CES | 326K | 840 | 1 GB DDR3 + 128 MB Linear BPI flash + 128 Mb Quad SPI flash |
| **Xilinx Virtex-7 AMD VC709** [107] | Virtex-7 XC7VX485T-2FFG1761C | 485K | 3.6K | DDR3 SODIMM + BPI Parallel NOR Flash: 32MB + IIC EEPROM:1KB (8Kb) |
| **Intel Altera Cyclone V (terasic DE-10 Development Kit)** [108] | Cyclone V | 110K | 112 | HPS SDRAM 1 GB DDR3 + 64MB FPGA SDRAM + EPCS128 Flash |
| **Intel Cyclone 10 GX FPGA Kit** [109] | Cyclone 10 | 220K | 192 | 1 channel of x40 DDR3 @ 933 MHz + EPCQ-L Flash + QSPI Flash |

These specialized devices offer a tailored solution for complex AI and deep learning tasks within AV systems, granting them high flexibility, high performance, and low power in hardware implementation [111]. Developed as a stand-alone device, the TPU is finely tuned for neural networks and is designed to work seamlessly with the Google TensorFlow framework [6]. This ASIC targets high volumes of low-precision arithmetic, particularly 8-bit calculations, and has already been leveraged across various applications at Google, including the search engine and AlphaGo [6].

The TPU v4 model comprises four chips, each with two cores as shown in Fig 15, and can compute more than 275 teraflops (BF16 or INT8) [112]. These cores incorporate scalar units, vector units, and 4 128x128 matrix units, all interconnected with on-chip 32GB high bandwidth memory (HBM) to facilitate pulsating matrix calculations. Notably, the TPU's performance is heightened by its ability to execute 16K multiply-accumulate operations in each cycle through one matrix unit per core employing BF16 precision. Moreover, other ASIC solutions, such as the EdgeTPU AI accelerator can achieve a remarkable 4 TOPS while consuming just 2 watts of power [113]. For instance, it can efficiently run cutting-edge mobile vision models like MobileNet V2 at nearly 400 FPS while conserving power [113].

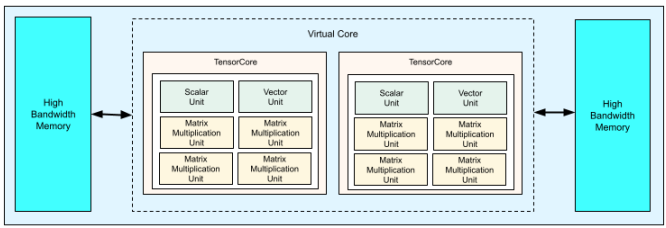

*Figure 15. TPU v4 Chip* [112]

A study conducted by [119] showed that Google's TPU v4 outperforms Nvidia A100 GPUs, demonstrating a 1.2 to 1.7 times faster speed, while simultaneously consuming 1.3 to 1.9 times less power than the Nvidia A100 GPU. In another study conducted by [111], a comparative analysis of ASICs with other hardware accelerators, including CPUs and GPUs, in the context of autonomous driving tasks unveiled several significant insights. Firstly, ASICs exhibit a substantial reduction in power consumption with almost a seven-fold improvement in energy efficiency for tasks like object detection. Additionally, when assessing power-hungry accelerators like GPUs, ASICs have the potential to significantly mitigate the thermal constraints, limiting the reduction in the vehicle's driving range to under 5%. Furthermore, ASIC-accelerated systems can markedly enhance the system's performance, reducing tail latency by a substantial factor, up to 93 times. This underscores their crucial role in maintaining consistent and responsive operations in AV systems when compared to GPUs, thereby ensuring reliability and safety in real-time applications. Notably, specialized ASICs like Google's TPU excel in lower-precision calculations, providing high throughput for training and inference in neural networks [120].

### 3) Heterogenous Hardware Platforms

Table IV offers an insight into the performance of different hardware implementations to run algorithms like SSD, CNN, YOLO, MobileNet, and others for object detection and classification. Each device is evaluated in terms of latency, accuracy, execution time, and power consumption. Notable findings include the diverse performance characteristics, with GPUs generally providing fast execution times but higher power consumption, while FPGAs and ASICs like the offer impressive accuracy with low power usage. These insights can be valuable for selecting the right hardware for specific algorithmic applications targeting AVs. Additionally, Table IV underscores the significance of heterogeneous hardware accelerators in modern computing. As the demands of various algorithms and datasets vary significantly, the availability of diverse hardware options is critical. Heterogeneous hardware accelerators enable organizations and researchers to tailor their hardware choices that align with their algorithmic and computational goals, ultimately leading to more efficient and effective implementations across a broad spectrum of applications.

TABLE IV
COMPARISON OF DIFFERENT HARDWARE IMPLEMENTATIONS ACROSS VARIOUS OBJECT DETECTION ALGORITHMS USED IN AVs

| Type | CPU | | | GPU | | | FPGA | | ASIC | |
|---|---|---|---|---|---|---|---|---|---|---|
| Source | [114] | [115] | [116] | [117] | | [118] | [115] | [116] | [117] | |
| Device | Intel i7-7700 | Intel core i7-4770 | NVIDIA GTX1060 | Nvidia Jetson Xavier | | XILINX ZYNQ ZCU102 | Xilinx ZC706 | Intel Arria 10 GX | Google Edge TPU | |
| Algorithm | CBFF-SSD | CNN for traffic sign detection | Speed-sign recognition algorithm | MobileNet V2 | Inception V3 | YOLOv2 | CNN for Stop-sign detection | Speed-sign recognition algorithm | MobileNet V2 | Inception V3 |
| Dataset | NWPU VHR-10 dataset | Real time video Input | LISA dataset | COCO dataset | COCO dataset | COCO dataset | Real time video input | LISA dataset | COCO dataset | COCO dataset |
| Latency | 382.15 | - | - | 2.57 | 14.51 | 5.376 | - | - | 3.5 | 52.77 |
| Accuracy | - | - | 92% | 71.15% | 77.82% | 76.21% | 99.8% | 92% | 70.94% | 77.62% |
| Execution Time | - | 136.2 ms | 30.3 ms | 24039 ms | 42808 ms | 0.244 s | 7.9 ms | 33.3 ms | 6051 ms | 17456 ms |
| Power Consumption | 65 W | 76 W | 19 W | 10.47 W | 21.84 W | 5.376 W | 5.2 W | 12.5 W | 4.89 W | 4.68 W |

## V. CONCLUSION

In conclusion, the evolving landscape of AVs demands a meticulous integration of hardware accelerators and sophisticated machine vision algorithms. This review paper has presented a comprehensive examination of different types of hardware accelerators and their features and sophisticated machine vision algorithms generally used for AVs, shedding light on the advancements in the field. The evolution of GPU-based hardware accelerators has been fundamental in addressing the computational demands of real-time processing for commercial autonomous vehicles. Simultaneously, the development of machine vision algorithms has been instrumental in enhancing the perception of autonomous vehicles. However, to meet the increasing computational demands of machine vision algorithms for autonomous vehicles, it is vital to consider other potential solutions such as FPGAs and TPUs and how can they be integrated into autonomous vehicles to offload some tasks from commercial hardware accelerators, paving the path for new heterogenous hardware solutions in autonomous vehicles. The synergy between hardware accelerators and machine vision algorithms has paved the way for advancements in autonomous vehicle technology. Looking into the future, the ongoing collaboration between researchers, engineers, and the industry will yield more robust hardware accelerators to meet the ever-increasing computational demands of machine vision algorithms to tackle the challenges faced in the field.

## VI. REFERENCE

[1] H. 'Field, "California DMV suspends Cruise's self-driving car permits, effective immediately," CNBC. Accessed: Nov. 16, 2023. [Online]. Available: https://www.cnbc.com/2023/10/24/california-dmv-suspends-cruises-self-driving-car-permits.html

[2] "Road traffic injuries," World Health Organization. Accessed: Oct. 15, 2023. [Online]. Available: https://www.who.int/news-room/fact-sheets/detail/road-traffic-injuries

[3] Wilson Kehoe Winingham staff, "Common Causes of Car Accidents," Wilson Kehoe Winingham. Accessed: Oct. 15, 2023. [Online]. Available: https://www.wkw.com/auto-accidents/blog/10-common-causes-traffic-accidents/

[4] C. Badue et al., "Self-driving cars: A survey," Expert Syst Appl, vol. 165, p. 113816, Mar. 2021, doi: 10.1016/j.eswa.2020.113816.

[5] S. Devi, P. Malarvezhi, R. Dayana, and K. Vadivukkarasi, "A Comprehensive Survey on Autonomous Driving Cars: A Perspective View," Wirel Pers Commun, vol. 114, no. 3, pp. 2121–2133, Oct. 2020, doi: 10.1007/s11277-020-07468-y.

[6] X. Feng, Y. Jiang, X. Yang, M. Du, and X. Li, "Computer vision algorithms and hardware implementations: A survey," Integration, vol. 69, pp. 309–320, Nov. 2019, doi: 10.1016/j.vlsi.2019.07.005.

[7] D. J. Yeong, G. Velasco-Hernandez, J. Barry, and J. Walsh, "Sensor and Sensor Fusion Technology in Autonomous Vehicles: A Review," Sensors, vol. 21, no. 6, p. 2140, Mar. 2021, doi: 10.3390/s21062140.

[8] M. Hasanujjaman, M. Z. Chowdhury, and Y. M. Jang, "Sensor Fusion in Autonomous Vehicle with Traffic Surveillance Camera System: Detection, Localization, and AI Networking," Sensors, vol. 23, no. 6, p. 3335, Mar. 2023, doi: 10.3390/s23063335.

[9] J. Vargas, S. Alsweiss, O. Toker, R. Razdan, and J. Santos, "An Overview of Autonomous Vehicles Sensors and Their Vulnerability to Weather Conditions," Sensors, vol. 21, no. 16, p. 5397, Aug. 2021, doi: 10.3390/s21165397.

[10] R. 'Digiuseppe, "Enabling Integrated ADAS Domain Controllers With Automotive IP," Semiconductor Engineering. Accessed: Oct. 23, 2023. [Online]. Available: https://semiengineering.com/enabling-integrated-adas-domain-controllers-with-automotive-ip/

[11] F. 'Narisawa et al., "Vehicle Electronic Control Units for Autonomous Driving in Safety and Comfort," Hitachi Review. Accessed: Oct. 23, 2023. [Online]. Available: https://www.hitachi.com/rev/archive/2022/r2022_01/01c01/index.html

[12] C. Wang, X. Wang, H. Hu, Y. Liang, and G. Shen, "On the Application of Cameras Used in Autonomous Vehicles," Archives of Computational Methods in Engineering, vol. 29, no. 6, pp. 4319–4339, Oct. 2022, doi: 10.1007/s11831-022-09741-8.

[13] J. 'Shepard, "How many types of radar are there?," Sensor Tips. Accessed: Oct. 23, 2023. [Online]. Available: https://www.sensortips.com/featured/how-many-types-of-radar-are-there-faq/

[14] P. Wei, L. Cagle, T. Reza, J. Ball, and J. Gafford, "LiDAR and Camera Detection Fusion in a Real-Time Industrial Multi-Sensor Collision Avoidance System," Electronics (Basel), vol. 7, no. 6, p. 84, May 2018, doi: 10.3390/electronics7060084.

[15] B. 'Dickson, "Is camera-only the future of self-driving cars?," ADAS & Autonomous Vehicle International. Accessed: Oct. 23, 2023. [Online]. Available: https://www.autonomousvehicleinternational.com/features/is-camera-only-the-future-of-self-driving-cars.html

[16] T. 'Thadani, R. 'Lerman, I. 'Piper, F. 'Siddiqui, and I. 'Uraizee, "The final 11 seconds of a fatal Tesla Autopilot crash," The Washington Post. Accessed: Dec. 04, 2023. [Online]. Available: https://www.washingtonpost.com/technology/interactive/2023/tesla-autopilot-crash-analysis/

[17] K. 'Power, S. 'Deva, T. 'Wang, J. 'Li, and C. 'Eising, "Hardware Accelerators in Autonomous Driving," in Proceedings of the Irish Machine Vision and Image Processing Conference 2023, USA, Aug. 2023.

[18] G. 'Sanders, "Autonomous Vehicle Sensors - Making Sense of The World," Wards Intelligence. Accessed: Oct. 23, 2023. [Online]. Available: https://wardsintelligence.informa.com/WI965823/Autonomous-Vehicle-Sensors---Making-Sense-of-The-World

[19] M. 'Frąckiewicz, "The Role of AI Hardware Accelerators in Autonomous Vehicles," TS2 SPACE. Accessed: Oct. 23, 2023. [Online]. Available: https://ts2.space/en/the-role-of-ai-hardware-accelerators-in-autonomous-vehicles/

[20] V. Alonso, A. Dacal-Nieto, L. Barreto, A. Amaral, and E. Rivero, "Industry 4.0 implications in machine vision metrology: an overview," Procedia Manuf, vol. 41, pp. 359–366, 2019, doi: 10.1016/j.promfg.2019.09.020.

[21] I. Sonata, Y. Heryadi, L. Lukas, and A. Wibowo, "Autonomous car using CNN deep learning algorithm," J Phys Conf Ser, vol. 1869, no. 1, p. 012071, Apr. 2021, doi: 10.1088/1742-6596/1869/1/012071.

[22] K. Othman, "Exploring the implications of autonomous vehicles: a comprehensive review," Innovative Infrastructure Solutions, vol. 7, no. 2, p. 165, Apr. 2022, doi: 10.1007/s41062-022-00763-6.

[23] A. Balasubramaniam and S. Pasricha, "Object Detection in Autonomous Vehicles: Status and Open Challenges," Jan. 2022.

[24] "Model 3 Owner's Manual," Tesla. Accessed: Sep. 20, 2023. [Online]. Available: https://www.tesla.com/ownersmanual/model3/en_jo/GUID-682FF4A7-D083-4C95-925A-5EE3752F4865.html

[25] "Volvo Cars to launch UK's largest and most ambitious autonomous driving trial," Volvo Cars Global Media Newsroom. Accessed: Oct. 15, 2023. [Online]. Available: https://www.media.volvocars.com/global/en-gb/media/pressreleases/189969/volvo-cars-to-launch-uks-largest-and-most-ambitious-autonomous-driving-trial

[26] S. Singh and B. S. Saini, "Autonomous cars: Recent developments, challenges, and possible solutions," IOP Conf Ser Mater Sci Eng, vol. 1022, no. 1, p. 012028, Jan. 2021, doi: 10.1088/1757-899X/1022/1/012028.

[27] P. 'Valdes, "GM self-driving car subsidiary withheld video of a crash, California DMV says," CNN Business. Accessed: Oct. 28, 2023. [Online]. Available: https://edition.cnn.com/2023/10/24/business/california-dmv-cruise-permit-revoke/index.html

[28] S. Casas, W. Luo, and R. Urtasun, "IntentNet: Learning to Predict Intention from Raw Sensor Data," 2018.

[29] W. Luo, B. Yang, and R. Urtasun, "Fast and Furious: Real Time End-to-End 3D Detection, Tracking and Motion Forecasting with a Single Convolutional Net," Dec. 2020, [Online]. Available: http://arxiv.org/abs/2012.12395

[30] E. Xie *et al.*, "DetCo: Unsupervised Contrastive Learning for Object Detection," in *2021 IEEE/CVF International Conference on Computer Vision (ICCV)*, IEEE, Oct. 2021, pp. 8372–8381. doi: 10.1109/ICCV48922.2021.00828.

[31] Z. Dai, B. Cai, Y. Lin, and J. Chen, "UP-DETR: Unsupervised Pre-training for Object Detection with Transformers," in *2021 IEEE/CVF Conference on Computer Vision and Pattern Recognition (CVPR)*, IEEE, Jun. 2021, pp. 1601–1610. doi: 10.1109/CVPR46437.2021.00165.

[32] A. Bar *et al.*, "DETReg: Unsupervised Pretraining with Region Priors for Object Detection," in *2022 IEEE/CVF Conference on Computer Vision and Pattern Recognition (CVPR)*, IEEE, Jun. 2022, pp. 14585–14595. doi: 10.1109/CVPR52688.2022.01420.

[33] P. Sun *et al.*, "Scalability in Perception for Autonomous Driving: Waymo Open Dataset," in *2020 IEEE/CVF Conference on Computer Vision and Pattern Recognition (CVPR)*, IEEE, Jun. 2020, pp. 2443–2451. doi: 10.1109/CVPR42600.2020.00252.

[34] T. Turay and T. Vladimirova, "Toward Performing Image Classification and Object Detection With Convolutional Neural Networks in Autonomous Driving Systems: A Survey," *IEEE Access*, vol. 10, pp. 14076–14119, 2022, doi: 10.1109/ACCESS.2022.3147495.

[35] Y. Cao, C. Li, Y. Peng, and H. Ru, "MCS-YOLO: A Multiscale Object Detection Method for Autonomous Driving Road Environment Recognition," *IEEE Access*, vol. 11, pp. 22342–22354, 2023, doi: 10.1109/ACCESS.2023.3252021.

[36] D. Parekh *et al.*, "A Review on Autonomous Vehicles: Progress, Methods and Challenges," *Electronics (Basel)*, vol. 11, no. 14, p. 2162, Jul. 2022, doi: 10.3390/electronics11142162.

[37] Y. Wang, H. Wang, and Z. Xin, "Efficient Detection Model of Steel Strip Surface Defects Based on YOLO-V7," *IEEE Access*, vol. 10, pp. 133936–133944, 2022, doi: 10.1109/ACCESS.2022.3230894.

[38] H. Slimani, J. El Mhamdi, and A. Jilbab, "Artificial Intelligence-based Detection of Fava Bean Rust Disease in Agricultural Settings: An Innovative Approach," *International Journal of Advanced Computer Science and Applications*, vol. 14, no. 6, 2023, doi: 10.14569/IJACSA.2023.0140614.

[39] J. Terven and D. Cordova-Esparza, "A Comprehensive Review of YOLO: From YOLOv1 and Beyond," Apr. 2023, [Online]. Available: http://arxiv.org/abs/2304.00501

[40] L. Liu, C. Ke, H. Lin, and H. Xu, "Research on Pedestrian Detection Algorithm Based on MobileNet-YoLo," *Comput Intell Neurosci*, vol. 2022, pp. 1–12, Oct. 2022, doi: 10.1155/2022/8924027.

[41] X. Zhai, Z. Huang, T. Li, H. Liu, and S. Wang, "YOLO-Drone: An Optimized YOLOv8 Network for Tiny UAV Object Detection," *Electronics (Basel)*, vol. 12, no. 17, p. 3664, Aug. 2023, doi: 10.3390/electronics12173664.

[42] J. Kim, J.-Y. Sung, and S. Park, "Comparison of Faster-RCNN, YOLO, and SSD for Real-Time Vehicle Type Recognition," in *2020 IEEE International Conference on Consumer Electronics - Asia (ICCE-Asia)*, IEEE, Nov. 2020, pp. 1–4. doi: 10.1109/ICCE-Asia49877.2020.9277040.

[43] S. A. Sanchez, H. J. Romero, and A. D. Morales, "A review: Comparison of performance metrics of pretrained models for object detection using the TensorFlow framework," *IOP Conf Ser Mater Sci Eng*, vol. 844, p. 012024, Jun. 2020, doi: 10.1088/1757-899X/844/1/012024.

[44] Y. Xing *et al.*, "Advances in Vision-Based Lane Detection: Algorithms, Integration, Assessment, and Perspectives on ACP-Based Parallel Vision," *IEEE/CAA Journal of Automatica Sinica*, vol. 5, no. 3, pp. 645–661, May 2018, doi: 10.1109/JAS.2018.7511063.

[45] M. Saranya, N. Archana, M. Janani, and R. Keerthishree, "Lane Detection in Autonomous Vehicles Using AI," 2023, pp. 15–30. doi: 10.1007/978-3-031-38669-5_2.

[46] W. Hao, "Review on lane detection and related methods," *Cognitive Robotics*, vol. 3, pp. 135–141, 2023, doi: 10.1016/j.cogr.2023.05.004.

[47] Bei He, Rui Ai, Yang Yan, and Xianpeng Lang, "Lane marking detection based on Convolution Neural Network from point clouds," in *2016 IEEE 19th International Conference on Intelligent Transportation Systems (ITSC)*, IEEE, Nov. 2016, pp. 2475–2480. doi: 10.1109/ITSC.2016.7795954.

[48] D. Tian, Y. Han, B. Wang, T. Guan, and W. Wei, "A Review of Intelligent Driving Pedestrian Detection Based on Deep Learning," *Comput Intell Neurosci*, vol. 2021, pp. 1–16, Jul. 2021, doi: 10.1155/2021/5410049.

[49] P. Felzenszwalb, D. McAllester, and D. Ramanan, "A discriminatively trained, multiscale, deformable part model," in *2008 IEEE Conference on Computer Vision and Pattern Recognition*, IEEE, Jun. 2008, pp. 1–8. doi: 10.1109/CVPR.2008.4587597.

[50] Y. Xiao *et al.*, "Deep learning for occluded and multi-scale pedestrian detection: A review," *IET Image Process*, vol. 15, no. 2, pp. 286–301, Feb. 2021, doi: 10.1049/ipr2.12042.

[51] W. Liu, S. Liao, W. Hu, X. Liang, and X. Chen, "Learning Efficient Single-Stage Pedestrian Detectors by Asymptotic Localization Fitting," 2018, pp. 643–659. doi: 10.1007/978-3-030-01264-9_38.

[52] S. K. Divvala, D. Hoiem, J. H. Hays, A. A. Efros, and M. Hebert, "An empirical study of context in object detection," in *2009 IEEE Conference on Computer Vision and Pattern Recognition*, IEEE, Jun. 2009, pp. 1271–1278. doi: 10.1109/CVPR.2009.5206532.

[53] Z. Cai, M. Saberian, and N. Vasconcelos, "Learning Complexity-Aware Cascades for Deep Pedestrian Detection," in *2015 IEEE International Conference on Computer Vision (ICCV)*, IEEE, Dec. 2015, pp. 3361–3369. doi: 10.1109/ICCV.2015.384.

[54] W. Lan, J. Dang, Y. Wang, and S. Wang, "Pedestrian Detection Based on YOLO Network Model," in *2018 IEEE International Conference on Mechatronics and Automation (ICMA)*, IEEE, Aug. 2018, pp. 1547–1551. doi: 10.1109/ICMA.2018.8484698.

[55] J. Wang, H. Li, S. Yin, and Y. Sun, "Research on Improved Pedestrian Detection Algorithm Based on Convolutional Neural Network," in *2019 International Conference on Internet of Things (iThings) and IEEE Green Computing and Communications (GreenCom) and IEEE Cyber, Physical and Social Computing (CPSCom) and IEEE Smart Data (SmartData)*, IEEE, Jul. 2019, pp. 254–258. doi: 10.1109/iThings/GreenCom/CPSCom/SmartData.2019.00063.

[56] R. Ayachi, M. Afif, Y. Said, and A. Ben Abdelaali, "pedestrian detection for advanced driving assisting system: a transfer learning approach," in *2020 5th International Conference on Advanced Technologies for Signal and Image Processing (ATSIP)*, IEEE, Sep. 2020, pp. 1–5. doi: 10.1109/ATSIP49331.2020.9231559.

[57] D. Tian, Y. Han, B. Wang, T. Guan, and W. Wei, "A Review of Intelligent Driving Pedestrian Detection Based on Deep Learning," *Comput Intell Neurosci*, vol. 2021, pp. 1–16, Jul. 2021, doi: 10.1155/2021/5410049.

[58] N. Triki, M. Karray, and M. Ksantini, "A Real-Time Traffic Sign Recognition Method Using a New Attention-Based Deep Convolutional Neural Network for Smart Vehicles," *Applied Sciences*, vol. 13, no. 8, p. 4793, Apr. 2023, doi: 10.3390/app13084793.

[59] Y. Wei, M. Gao, J. Xiao, C. Liu, Y. Tian, and Y. He, "Research and Implementation of Traffic Sign Recognition Algorithm Model Based on Machine Learning," *Journal of Software Engineering and Applications*, vol. 16, no. 06, pp. 193–210, 2023, doi: 10.4236/jsea.2023.166011.

[60] Y. Li, J. Li, and P. Meng, "Attention-YOLOV4: a real-time and high-accurate traffic sign detection algorithm," *Multimed Tools Appl*, vol. 82, no. 5, pp. 7567–7582, Feb. 2023, doi: 10.1007/s11042-022-13251-x.

[61] T. P. Dang, N. T. Tran, V. H. To, and M. K. Tran Thi, "Improved YOLOv5 for real-time traffic signs recognition in bad weather conditions," *J Supercomput*, vol. 79, no. 10, pp. 10706–10724, Jul. 2023, doi: 10.1007/s11227-023-05097-3.

[62] R. Kulkarni, S. Dhavalikar, and S. Bangar, "Traffic Light Detection and Recognition for Self Driving Cars Using Deep Learning," in *2018 Fourth International Conference on Computing Communication Control and Automation (ICCUBEA)*, IEEE, Aug. 2018, pp. 1–4. doi: 10.1109/ICCUBEA.2018.8697819.

[63] M. Takaki and H. Fujiyoshi, "Traffic Sign Recognition Using SIFT Features," *IEEJ Transactions on Electronics, Information and Systems*, vol. 129, no. 5, pp. 824–831, 2009, doi: 10.1541/ieejeiss.129.824.

[64] Md. Z. Abedin, P. Dhar, and K. Deb, "Traffic Sign Recognition using SURF: Speeded up robust feature descriptor and artificial

neural network classifier," in *2016 9th International Conference on Electrical and Computer Engineering (ICECE)*, IEEE, Dec. 2016, pp. 198–201. doi: 10.1109/ICECE.2016.7853890.

[65] H. Gao, C. Liu, Y. Yu, and B. Li, "Traffic signs recognition based on PCA-SIFT," in *Proceeding of the 11th World Congress on Intelligent Control and Automation*, IEEE, Jun. 2014, pp. 5070–5076. doi: 10.1109/WCICA.2014.7053576.

[66] X.-H. Wu, R. Hu, and Y.-Q. Bao, "Pedestrian traffic light detection in complex scene using AdaBoost with multi-layer features," 2018.

[67] Z. Ozcelik, C. Tastimur, M. Karakose, and E. Akin, "A vision based traffic light detection and recognition approach for intelligent vehicles," in *2017 International Conference on Computer Science and Engineering (UBMK)*, IEEE, Oct. 2017, pp. 424–429. doi: 10.1109/UBMK.2017.8093430.

[68] A. Vinod Deshpande and A. V Deshpande Assistant Professor, "Design Approach for a Novel Traffic Sign Recognition System by Using LDA and Image Segmentation by Exploring the Color and Shape Features of an Image," 2014. [Online]. Available: https://www.researchgate.net/publication/342674166

[69] M. DeRong and T. ZhongMei, "Remote Traffic Light Detection and Recognition Based on Deep Learning," in *2023 6th World Conference on Computing and Communication Technologies (WCCCT)*, IEEE, Jan. 2023, pp. 194–198. doi: 10.1109/WCCCT56755.2023.10052610.

[70] P. Liu and T. Li, "Traffic light detection based on depth improved YOLOV5," in *2023 3rd International Conference on Neural Networks, Information and Communication Engineering (NNICE)*, IEEE, Feb. 2023, pp. 395–399. doi: 10.1109/NNICE58320.2023.10105786.

[71] N. H. Sarhan and A. Y. Al-Omary, "Traffic light Detection using OpenCV and YOLO," in *2022 International Conference on Innovation and Intelligence for Informatics, Computing, and Technologies (3ICT)*, IEEE, Nov. 2022, pp. 604–608. doi: 10.1109/3ICT56508.2022.9990723.

[72] H. T. Ngoc, K. H. Nguyen, H. K. Hua, H. V. N. Nguyen, and L.-D. Quach, "Optimizing YOLO Performance for Traffic Light Detection and End-to-End Steering Control for Autonomous Vehicles in Gazebo-ROS2," *International Journal of Advanced Computer Science and Applications*, vol. 14, no. 7, 2023, doi: 10.14569/IJACSA.2023.0140752.

[73] S. Pavlitska, N. Lambing, A. K. Bangaru, and J. M. Zöllner, "Traffic Light Recognition using Convolutional Neural Networks: A Survey," Sep. 2023, [Online]. Available: http://arxiv.org/abs/2309.02158

[74] S.-Y. Lin and H.-Y. Lin, "A Two-Stage Framework for Diverse Traffic Light Recognition Based on Individual Signal Detection," 2022, pp. 265–278. doi: 10.1007/978-3-031-04112-9_20.

[75] "LISA Traffic Light Dataset," Kaggle. Accessed: Dec. 06, 2023. [Online]. Available: https://www.kaggle.com/datasets/mbornoe/lisa-traffic-light-dataset

[76] "Bosch Small Traffic Lights Dataset," Heidelberg Collaboratory for Image Processing (HCI). Accessed: Dec. 06, 2023. [Online]. Available: https://hci.iwr.uni-heidelberg.de/content/bosch-small-traffic-lights-dataset

[77] "DriveU Traffic Light Dataset (DTLD)," UULM. Accessed: Dec. 06, 2023. [Online]. Available: https://www.uni-ulm.de/in/iui-drive-u/projekte/driveu-traffic-light-dataset/

[78] Steve, "Tesla Hardware 3 (Full Self-Driving Computer) Detailed," AutoPilot Review. Accessed: Sep. 22, 2023. [Online]. Available: https://www.autopilotreview.com/tesla-custom-ai-chips-hardware-3/

[79] "FSD chip - tesla," WikiChip. Accessed: Sep. 22, 2023. [Online]. Available: https://en.wikichip.org/wiki/tesla_(car_company)/fsd_chip

[80] C. 'Bos, "Tesla's New HW3 Self-Driving Computer - It's A Beast (CleanTechnica Deep Dive)," CleanTechnica. Accessed: Sep. 22, 2023. [Online]. Available: https://cleantechnica.com/2019/06/15/teslas-new-hw3-self-driving-computer-its-a-beast-cleantechnica-deep-dive/

[81] A. 'Mishra, "Decoding the Technology Behind Tesla Autopilot: How it Works," Medium. Accessed: Nov. 26, 2023. [Online]. Available: https://ai.plainenglish.io/decoding-the-technology-behind-tesla-autopilot-how-it-works-af92cdd5605f

[82] "Jetson - Embedded AI Computing Platform," NVIDIA Developer. Accessed: Nov. 24, 2023. [Online]. Available: https://developer.nvidia.com/embedded-computing

[83] "NVIDIA Jetson Orin," NVIDIA. Accessed: Nov. 25, 2023. [Online]. Available: https://www.nvidia.com/en-us/autonomous-machines/embedded-systems/jetson-orin/

[84] Y. Kortli, S. Gabsi, L. F. C. L. Y. Voon, M. Jridi, M. Merzougui, and M. Atri, "Deep embedded hybrid CNN–LSTM network for lane detection on NVIDIA Jetson Xavier NX," *Knowl Based Syst*, vol. 240, p. 107941, Mar. 2022, doi: 10.1016/j.knosys.2021.107941.

[85] B. R. Chang, H.-F. Tsai, and C.-W. Hsieh, "Accelerating the Response of Self-Driving Control by Using Rapid Object Detection and Steering Angle Prediction," *Electronics (Basel)*, vol. 12, no. 10, p. 2161, May 2023, doi: 10.3390/electronics12102161.

[86] " Mercedes-Benz and Nvidia to build autonomous driving software," Autovista24. Accessed: Sep. 23, 2023. [Online]. Available: https://autovista24.autovistagroup.com/news/mercedes-benz-and-nvidia-build-autonomous-driving-software/

[87] " Volvo Cars deepens collaboration with NVIDIA; next-generation self-driving Volvos powered by NVIDIA DRIVE Orin," Volvo Cars Global Media Newsroom. Accessed: Sep. 23, 2023. [Online]. Available: https://www.media.volvocars.com/global/en-gb/media/pressreleases/280495/volvo-cars-deepens-collaboration-with-nvidia-next-generation-self-driving-volvos-powered-by-nvidia-d

[88] M. 'Hibben, "Nvidia: Technology Leadership In Advanced Driver Assistance (NASDAQ:NVDA)," Seeking Alpha. Accessed: Sep. 23, 2023. [Online]. Available: https://seekingalpha.com/article/4530275-nvidia-technology-leadership-in-advanced-driver-assistance

[89] "AUTONOMOUS DRIVING The future of ADAS and automated driving has arrived," Qualcomm. Accessed: Sep. 20, 2023. [Online]. Available: https://www.qualcomm.com/products/automotive/autonomous-driving#

[90] P. 'Moorhead, "Qualcomm Officially Enters Self-Driving Market With Snapdragon Ride Platform And Extends Partnership With GM To Include ADAS," Moor Insights & Strategy. Accessed: Sep. 19, 2023. [Online]. Available: https://moorinsightsstrategy.com/qualcomm-officially-enters-self-driving-market-with-snapdragon-ride-platform-and-extends-partnership-with-gm-to-include-adas/

[91] "Snapdragon Ride Platform," Qualcomm Developer Network. Accessed: Sep. 19, 2023. [Online]. Available: https://developer.qualcomm.com/software/digital-chassis/snapdragon-ride/snapdragon-ride-platform#

[92] C. 'Hammerschmidt, "BMW picks Qualcomm for Automated Driving systems," eeNews Europe. Accessed: Sep. 19, 2023. [Online]. Available: https://www.eenewseurope.com/en/bmw-picks-qualcomm-for-automated-driving-systems/

[93] S. 'Abuelsamid, "Mobileye Announces EyeQ6 And EyeQ Ultra Chips For Assisted And Automated Driving," Forbes. Accessed: Nov. 28, 2023. [Online]. Available: https://www.forbes.com/sites/samabuelsamid/2022/01/04/mobileye-announces-eyeq6-and-eyeq-ultra-chips-for-assisted-and-automated-driving/?sh=7730f20b79e6

[94] "Mobileye EyeQ6 series in-depth analysis," inf.news. Accessed: Dec. 05, 2023. [Online]. Available: https://inf.news/en/tech/6176daaf60e89f0febe1fdec75e2ea47.html

[95] "New Mobileye EyeQ Ultra will Enable Consumer AVs," Mobileye. Accessed: Dec. 01, 2023. [Online]. Available: https://www.mobileye.com/news/mobileye-ces-2022-tech-news/

[96] "Intelligent Speed Assist Shows the Power of Mobileye's Vision," Mobileye. Accessed: Dec. 01, 2023. [Online]. Available: https://www.mobileye.com/blog/intelligent-speed-assist-isa-computer-vision-adas-solution/

[97] "Rethinking technology for the autonomous future," Mobileye. Accessed: Nov. 29, 2023. [Online]. Available: https://www.mobileye.com/technology/

[98] K. 'Korosec, "With Intel Mobileye's newest chip, automakers can bring automated driving to cars," TechCrunch. Accessed: Dec. 06, 2023. [Online]. Available:

https://techcrunch.com/2022/01/04/intels-mobileye-autonomous-driving-chip-for-consumer-vehicles/
[99] Z. 'Shahan, "Mobileye's Partnerships With BMW, Ford, NIO, Nissan, Volkswagen, & WILLER," CleanTechnica. Accessed: Dec. 06, 2023. [Online]. Available: https://cleantechnica.com/2020/08/10/mobileyes-partnerships-with-bmw-ford-nio-nissan-volkswagen-willer/
[100] R. 'Niranjana, "FPGAs in Self-Driving Cars: Accelerating Perception and Decision-Making," FPGA Insights. Accessed: Sep. 25, 2023. [Online]. Available: https://fpgainsights.com/fpga/fpgas-in-self-driving-cars-accelerating-perception-and-decision-making/
[101] R. Raj and R. Prakash, "FPGA Based Lane Tracking system for Autonomous Vehicles," 2019.
[102] "Zynq 7000 SoC," AMD. Accessed: Nov. 05, 2023. [Online]. Available: https://www.xilinx.com/products/silicon-devices/soc/zynq-7000.html
[103] "Kria KV260 Vision AI Starter Kit," AMD. Accessed: Nov. 05, 2023. [Online]. Available: https://www.xilinx.com/products/som/kria/kv260-vision-starter-kit.html
[104] "ALINX SoM AC7020C: SoC Industrial Grade Module," AMD. Accessed: Dec. 07, 2023. [Online]. Available: https://www.xilinx.com/products/boards-and-kits/1-1bkpidx.html
[105] "Ultra96-V2 | Avnet Boards," AVNET. Accessed: Nov. 05, 2023. [Online]. Available: https://www.avnet.com/wps/portal/us/products/avnet-boards/avnet-board-families/ultra96-v2/
[106] "AMD Kintex7 FPGA KC705 Evaluation Kit," AMD. Accessed: Nov. 05, 2023. [Online]. Available: https://www.xilinx.com/products/boards-and-kits/ek-k7-kc705-g.html
[107] " AMD Virtex 7 FPGA VC709 Connectivity Kit," AMD. Accessed: Nov. 05, 2023. [Online]. Available: https://www.xilinx.com/products/boards-and-kits/dk-v7-vc709-g.html
[108] "All FPGA Boards - Cyclone V - DE10-Standard," Terasic. Accessed: Nov. 05, 2023. [Online]. Available: https://www.terasic.com.tw/cgi-bin/page/archive.pl?Language=English&CategoryNo=167&No=1081
[109] "Intel® Cyclone® 10 GX Development Kit," Intel. Accessed: Nov. 05, 2023. [Online]. Available: https://www.intel.com/content/www/us/en/products/details/fpga/development-kits/cyclone/10-gx.html
[110] K. Shi, M. Wang, X. Tan, Q. Li, and T. Lei, "Efficient Dynamic Reconfigurable CNN Accelerator for Edge Intelligence Computing on FPGA," *Information*, vol. 14, no. 3, p. 194, Mar. 2023, doi: 10.3390/info14030194.
[111] S.-C. Lin *et al.*, "The Architectural Implications of Autonomous Driving: Constraints and Acceleration," in *Proceedings of the Twenty-Third International Conference on Architectural Support for Programming Languages and Operating Systems*, New York, NY, USA: ACM, Mar. 2018, pp. 751–766. doi: 10.1145/3173162.3173191.
[112] " System Architecture | Cloud TPU | Google Cloud," Google Cloud. Accessed: Sep. 26, 2023. [Online]. Available: https://cloud.google.com/tpu/docs/system-architecture-tpu-vm
[113] "Introduction to Cloud TPU | Google Cloud," Google Cloud. Accessed: Sep. 26, 2023. [Online]. Available: https://cloud.google.com/tpu/docs/intro-to-tpu/
[114] Li, Zhang, and Wu, "Efficient Object Detection Framework and Hardware Architecture for Remote Sensing Images," *Remote Sens (Basel)*, vol. 11, no. 20, p. 2376, Oct. 2019, doi: 10.3390/rs11202376.
[115] W. Shi, X. Li, Z. Yu, and G. Overett, "An FPGA-Based Hardware Accelerator for Traffic Sign Detection," *IEEE Trans Very Large Scale Integr VLSI Syst*, vol. 25, no. 4, pp. 1362–1372, Apr. 2017, doi: 10.1109/TVLSI.2016.2631428.
[116] M. Yih, J. M. Ota, J. D. Owens, and P. Muyan-Ozcelik, "FPGA versus GPU for Speed-Limit-Sign Recognition," in *2018 21st International Conference on Intelligent Transportation Systems (ITSC)*, IEEE, Nov. 2018, pp. 843–850. doi: 10.1109/ITSC.2018.8569462.
[117] H. Kong *et al.*, "EDLAB: A Benchmark for Edge Deep Learning Accelerators," *IEEE Des Test*, vol. 39, no. 3, pp. 8–17, Jun. 2022, doi: 10.1109/MDAT.2021.3095215.
[118] S. N. Tesema and E.-B. Bourennane, "Resource- and Power-Efficient High-Performance Object Detection Inference Acceleration Using FPGA," *Electronics (Basel)*, vol. 11, no. 12, p. 1827, Jun. 2022, doi: 10.3390/electronics11121827.
[119] N. Jouppi *et al.*, "TPU v4: An Optically Reconfigurable Supercomputer for Machine Learning with Hardware Support for Embeddings," in *Proceedings of the 50th Annual International Symposium on Computer Architecture*, New York, NY, USA: ACM, Jun. 2023, pp. 1–14. doi: 10.1145/3579371.3589350.
[120] J. 'Tovar, "GPUs vs TPUs: A Comprehensive Comparison for Neural Network Workloads," LinkedIn. Accessed: Sep. 22, 2023. [Online]. Available: https://www.linkedin.com/pulse/gpus-vs-tpus-comprehensive-comparison-neural-network-workloads-joel/
[121] I. Elmanaa, M. A. Sabri, Y. Abouch, and A. Aarab, "Efficient Roundabout Supervision: Real-Time Vehicle Detection and Tracking on Nvidia Jetson Nano," *Applied Sciences*, vol. 13, no. 13, p. 7416, Jun. 2023, doi: 10.3390/app13137416.
[122] L. S. Karumbunathan, "NVIDIA Jetson AGX Orin Series A Giant Leap Forward for Robotics and Edge AI Applications," 2022.
[123] "The Snapdragon Ride Platform continues to push ADAS/AD forward," Qualcomm. Accessed: Sep. 19, 2023. [Online]. Available: https://www.qualcomm.com/news/onq/2023/01/snapdragon-ride-platform-continues-to-push-adas-ad-forward